\documentclass[aps,prb,twocolumn,groupedaddress,showpacs,showkeys]{revtex4}
\usepackage{graphicx}
\begin{document}
\title{Numerical Solution of the 1D Schr\"odinger Equation: Bloch Wavefunctions }
\author{Constantino A. Utreras D\'{\i}az }
\email{cutreras@uach.cl} \affiliation{  Instituto de F\'{\i}sica, 
Facultad de Ciencias, Universidad Austral de Chile, 
Campus Isla Teja s/n, Casilla 567, Valdivia, Chile}
\date{\today}
\begin{abstract}
In this article we discuss a procedure to solve the one
dimensional (1D) Schr\"odinger Equation for a periodic potential,
which may be well suited to teach band structure theory. The
procedure is conceptually very simple, so that it may be used to
teach band theory at the undergraduate level; at the same time the
point of view is practical, so that the students may experiment
computing band gaps, and other features of band structure. Another
advantage of the procedure lies in that it does not use specific
symmetry properties of the potential, so that the results are
generally valid.
\end{abstract}
\maketitle

\section{Introduction}

Most traditional textbooks dealing with Solid State Theory at an
elementary level discuss the fundamentals of band structure
calculations, including in most cases only a theoretical
description of the essential features of band structure, usually
with a discussion of the Kronnig-Penney model~\cite{ASHCROFT}. Of
course, this is of great pedagogical value, but presenting only this viewpoint 
is unsatisfactory, given the great amount of computer resources available today, and the
relative ease with which  one may take advantage of the added insight that may be obtained from
simple numerical calculations.

Of course, this fact has not been neglected in scientific
literature; many articles have been devoted to the numerical solution of the Schr\"odinger 
equation in this and other journals~\cite{BAHURMUZ,SHAKIR,DECK,DENHAM,VDMAELEN,LEUNG, 
MENDEZ,GREENHOW,JOHNSTON,COONEY,SEARLES}. The references cited here intend only to be a
(biased) sample of the large existing literature on the subject of
numerically solving the Schr\"odinger equation for one dimensional
potentials. Most of the articles seem to concentrate on the
problem of finding bound states~\cite{SHAKIR,DENHAM,VDMAELEN,GREENHOW,COONEY,SEARLES},
while a relatively minor number treat the problem of finding the
energy spectra for a periodic potential~\cite{BAHURMUZ,DECK,LEUNG,MENDEZ,JOHNSTON}, 
which is the subject of this article.

The band theory of Bloch electrons in a periodic potential is
basic for an understanding of solid state physics.  The usual
description is given in most texts in solid state theory(see, for
example Aschcroft and Mermin's classic textbook~\cite{ASHCROFT}, 
which we  follow here for the one dimensional case.

Consider an electron (mass $m$) moving in a periodic potential
$U(x + a ) = U(x)$, in which the potential $U(x)$ is written as a
superposition of potential barriers $v(x)$, centered the points $x
= \pm (n+1/2) a$:
\begin{equation}
\label{Potencial} U(x) = \sum_{n=-\infty}^{\infty} v(x-na).
\end{equation}

The band structure of the one dimensional solid is expressed in
terms of the properties of an electron scattering from a single
potential barrier $v(x)$. Therefore, they~(\cite{ASHCROFT}) write
the wavefunction for an electron scattering from the left, with
energy $ E = \hbar^2 K^2/2m$, in terms of reflection and
transmission amplitudes $r$ and $t$; these wavefunction become,
for $x>a$

\begin{eqnarray}
\Psi_l(x) &=& e^{i Kx} + r e^{-i K x} , ~~ x<0 \nonumber \\ &=& t
e^{i K x}, ~~ x>a,
\end{eqnarray}
while, for the electron coming from the right-hand side, the
wavefunction is

\begin{eqnarray}
\Psi_r(x) &=& t e^{-i Kx} , ~~ x<0 \nonumber \\ &=& e^{-iKx} + r
e^{i K x}, ~~ x>a.
\end{eqnarray}
Observe that these results are valid only for a potential that is
even with respect to $x=a/2$, otherwise, one would have to
introduce different coefficients $r^{\prime}$ and $t^{\prime}$ in
the last equation. We assume, of course, that the rest of the
solutions, i.e., the part corresponding to the region $0 < x < a$
has been found by some procedure. Clearly, since $\Psi_l$ and
$\Psi_r$ are two independent solutions of the Schr\"odinger
equation corresponding to the same energy, the full solution of
the Schr\"odinger equation corresponding to the periodic solid
will be expressed as a the linear combination (now for $0<x<a$)

\begin{equation}
\Psi(x) = A \Psi_l (x) + B \Psi_r(x), ~~ 0<x<a.
\end{equation}

According to Bloch's Theorem, the wavefunction $\Psi(x)$ satisfies
the relation

\begin{equation}
\Psi(x + a ) =  e^{ika} \Psi(x),
\end{equation}
for suitable values of $k$. The same relations holds for the
derivative of $\Psi (x)$, namely

\begin{equation}
\Psi^{\prime}(x + a) = e^{ika}\Psi^{\prime}(x).
\end{equation}
Imposing the conditions above on the wavefunction $\Psi(x)$, gives
a relation that may be used to obtain the energy vs. wavevector
relation $ E = E(k) = \hbar^2 K^2/2m$,
\begin{equation}
\cos(ka) = \frac{t^2 - r^2}{2t} e^{iKa} + \frac{1}{2t}e^{-iKa}.
\end{equation}

It is then shown that the energy is a periodic function of the
Bloch wavenumber $k$, with period $G = 2 \pi / a$ (the reciprocal
lattice vector).

\begin{equation}
E_{k+G} = E_{k}
\end{equation}

Let us concentrate on the solution of the equation for the Bloch wavefunction
$u(x)$. One way to do this may be to expand the wavefunction $\Psi$ in a
Fourier series, doing the same for the potential, namely

\begin{eqnarray}
\Psi(x) &=& \sum_q c_q e^{iqx} \nonumber \\ U(x) &=& \sum_G U_G~
e^{iGx} \nonumber \\ U_G &=& \frac{1}{a} \int_{0}^{a} e^{-iGx}
U(x) dx \nonumber ,
\end{eqnarray}
in which the $U_G$ are Fourier coefficients, and the sum over $G$
goes over reciprocal lattice vectors.

The equation for the coefficients $c_q$ is customarily  written as follows
\newcommand{\Gp}{G^{\prime}}
\begin{equation}
\left( \frac{\hbar^2}{2m} (k-G)^2 - E \right) c_{k-G} + \sum_{\Gp}
U_{\Gp - G} c_{k - \Gp} = 0.
\end{equation}

From a numerical point of view, this equation may be slow in
converging to an accurate solution, one may need to include a
large number of terms in the resulting matrix equation. However,
this method has been successfully used to calculate band
structures of real solids (3D).

In the 1D case, we have other methods at our disposal, which may be better
suited to the problem. Unfortunately, most of the methods available for the
1D case cannot be taken over to the 3D case, and our proposal is not an
exception to this (alas!).

\section{Numerical Solution}

%
\newcommand{\Df}{a}
\newcommand{\Dc}{\Delta_C}
\newcommand{\Ds}{\Delta_S}
\newcommand{\Dcp}{\Dc^{\prime}}
\newcommand{\Dsp}{\Ds^{\prime}}
\newcommand{\Cp}{C^{\prime}}
\newcommand{\Sp}{S^{\prime}}

Now let us discuss the particular situation that arises in 1D. A peculiarity
of our treatment is that we really have no need for the auxiliary function
$u(x)$, as it will be seen presently.

Now, to actually find numerical solutions of the Schr\"odinger equation, and
the corresponding energies, we proceed as follows. For each value of the
energy $E$, considered here as a parameter, we find two independent numerical
solutions, which we call $C(x)$ and $S(x)$. For example, for a fundamental
period with endpoints at $x = 0$ and $x = a$, these functions may be chosen
to satisfy
\begin{eqnarray}
C(a/2) = 1 & \Cp (a/2) = 0 \\ S(a/2) = 0 & \Sp(a/2) = 1
\end{eqnarray}

To numerically obtain the functions $C$ and $S$ we divide the integration region
in two regions, $0< x < a/2$, and $a/2 < x < a$. We start the integration from the 
center ( $x=a/2$ towards the boundaries of the two regions, imposing at the start 
the continuity of the function ( $C$ os $S$). After this is done, we compute the 
normalization integral, $N = \int_0^a |\Psi(x)|^2 dx$ and redefine the function, 
dividing by $\sqrt{N}$. Observe that the wavefunction defined in this way, be it 
$C$ or $S$, does not satisfy the full boundary conditions: it is a real function, 
and it may not even be periodic, the boundary conditions are satisfied only by the
full wavefunction $\Psi(x)$. For the numerical integration, I have used the 
Numerov~\cite{CHOW} procedure, since it is very accurate and simple to use, however, 
any reliable numerical  method may be used just as easily, such as Runge-Kutta, with 
or without adaptive step size~\cite{RECIPES}.

The functions $C(x)$ and $S(x)$ are actually dependent upon the energy $E$,
i.e., $C(x) = C_E(x)$ and $S(x) = S_E (x)$, but to make the notation less
cumbersome, we drop the parameter $E$ henceforth, unless it is necessary 
to keep it. The full wavefunction must  be expressed as a the linear combination 
of the functions $C$ and $S$, with possibly complex coefficients $A$ and $B$, the 
only way to satisfy the boundary conditions from our purely real basis functions $C$ and $S$,

\begin{equation}
\Psi(x) = A \cdot C(x) + B\cdot S(x).
\end{equation}

The boundary conditions are now

\begin{eqnarray}
\Psi(a  ) = e^{ i k a } \Psi(0) \nonumber\\ \Psi^{\prime}(a) =
e^{i k a} \Psi^{\prime}(0),
\end{eqnarray}
which may be written as equations for the coefficients $A$ and
$B$,

\begin{eqnarray}
\Dc A + \Ds B & =  &  0 \nonumber \\ \Dcp A + \Dsp B & = & 0 ,
\end{eqnarray}
in which the coefficients are may be written as (using $\theta =
ka/2$)

\begin{eqnarray}
\Dc & = & e^{i \theta} C(a) - e^{-i\theta} C(0) \nonumber \\ \Ds &
= & e^{i \theta} S(a) - e^{-i \theta} S(0) \nonumber \\ \Dcp & = &
e^{i \theta} C^{\prime}(a) - e^{-i \theta} C^{\prime}(0) \nonumber
\\ \Dsp & = & e^{i \theta} S^{\prime}(a) - e^{-i \theta}
S^{\prime}(0)
\end{eqnarray}

The equation for the energies is then given by the vanishing of
the determinant of the coefficients,
\begin{equation}
G (E) = \Dc \Dsp - \Ds \Dcp = 0
\end{equation}
When, as in our case, the potential is symmetric with respect to
the midpoint of the period, the functions $C$ and $S$ have
definite parity about that point, i.e. $C(a) = C(0)$ and $S( a ) =
-S(0)$. Also, the derivatives have the opposing parity,
$C^{\prime}(a) = -C^{\prime}(0)$ and $S^{\prime}(a) =
S^{\prime}(0)$, therefore,

\begin{eqnarray}
\Dc & = & 2 i \sin( \theta) C(0) \nonumber \\ \Ds & = & -2\cos( \theta) S(0)
\nonumber\\ \Dcp & = & -2 \cos(\theta ) C^{\prime}(0) \nonumber
\\ \Dsp & = & 2 i \sin( \theta) S^{\prime}(0 )
\end{eqnarray}
The eigenvalue equation becomes

\begin{equation}
G(E) = \sin^2(\theta) C(0) S^{\prime}(0) + \cos^2( \theta) S(0)
C^{\prime}(0) = 0 \end{equation}

These equations are convenient for numerical work, since the functions $C$
and $S$ are easily computed numerically, and the left hand side of the
equation may be plotted, as a function of $E$, the zeros being the
eigenstates that we are looking for.

In the general (non symmetric) case, we can write

\begin{eqnarray}
\Dc & = & [ C(\Df) - C(0) ] \cos(\theta) + i [ C(\Df) + C(0)] \sin(\theta)
\nonumber \\ \Ds & = &  [S(\Df) - S(0)]\cos(\theta) + i [ S(\Df) + S(0)]
\sin(\theta) \nonumber \\ \Dcp & = & [ \Cp (\Df) - \Cp(0)] \cos(\theta) + i [
\Cp(\Df) + \Cp(0)] \sin(\theta) \nonumber \\ \Dsp & = & [ \Sp(\Df) - \Sp(0) ]
\cos(\theta) + i [ \Sp(\Df) + \Sp(0) ] \cos(\theta)
\end{eqnarray}

Now, let

\begin{eqnarray}
C_{(-)} & = & C(\Df) - C(0)\nonumber \\ C_{(+)} & = & C(\Df) + C(0) \nonumber
\\ S_{(-)} & = & S(\Df) - S(0) \nonumber \\ S_{(+)} & = &  S(\Df) + S(0) \nonumber \\ \Cp_{(-)}
& = &  \Cp (\Df) - \Cp(0) \nonumber \\ \Cp_{(+)} & = &  \Cp(\Df) + \Cp(0)
\nonumber \\ \Sp_{(-)} & = & \Sp(\Df) - \Sp(0)  \nonumber \\ \Sp_{(+)} & = &
\Sp(\Df) + \Sp(0) \nonumber .
\end{eqnarray}

We obtain, for the real and imaginary parts of the eigenvalue
equation:

\begin{eqnarray}
\Re [ G (E) ] & = & \left( C_{(-)} \Sp_{(-)} - S_{(-)} \Cp_{(-)}
\right) \cos^2(\theta) \nonumber
 \\ &+& \left( S_{(+)} \Cp_{(-)} -
C_{(+)} \Sp_{(-)}  \right) \sin^2(\theta) \nonumber
\\ \Im [ G(E) ] & = & \left(  C_{(-)} \Sp_{(+)}+ C_{(+)} \Sp_{(-)} - S_{(-)} \Cp_{(+)} - S_{(+)} \Cp_{(-)}
\right)\nonumber \\
&\times & \sin(\theta) \cos(\theta) \nonumber
\end{eqnarray}

First, concentrate upon computing the terms of the $\Im (G)$, they become

\begin{eqnarray}
 C_{(-)} \Sp_{(+)} + C_{(+)} \Sp_{(-)}& = &  2 [ C(\Df) \Sp(\Df) - C(0) \Sp(0) ] \nonumber \\
 S_{(-)} \Cp_{(+)} + S_{(+)} \Cp_{(-)} & = & 2 [ S(\Df) \Cp(\Df) - S(0) \Cp(0) ] \nonumber
\end{eqnarray}

Putting all this together, we find

\begin{equation}
\Im (G(E)) = 2 \sin(\theta)\cos(\theta) \left( W[C,S](\Df) - W[C,S](0)
\right) ,
\end{equation}
where $W [C,S](\phi = C(\phi) \Sp(\phi) -  \Cp(\phi) S(\phi)$ is the
Wronskian of the solutions $C$ and $S$; moreover, a well known theorem tells
us that the Wronskian is a constant, as it may easily shown, hence, we have
shown that the imaginary part of $G$ identically vanishes

\begin{equation}
\Im [ G(E) ] = 0.
\end{equation}

The eigenvalue equation is therefore,

\begin{eqnarray}
G (E) &=&  \left( C_{(-)} \Sp_{(-)} - S_{(-)} \Cp_{(-)} \right)
\cos^2(\theta) \nonumber \\ &+& \left( S_{(+)} \Cp_{(-)} - C_{(+)}
\Sp_{(-)} \right) \sin^2(\theta) \nonumber
\end{eqnarray}

\section{Simple Examples}

As indicated previously, a numerical calculation such as the one proposed here is 
able to give us all the energy bands that we care to compute (with due consideration 
of the limitations of each integration method). Since computing wavefunctions is relatively 
'cheap', we have developed a computer program that does the following:

\begin{enumerate}%
\item Divides the first Brillouin zone in $N$ intervals of size $\Delta k = 2 \pi/Na$.

\item Prompts the user to indicate the energy range, from $E_{min}= min( V(x)$, $0<x<a$) 
to $E_{max}$. The energy interval is divided into $N_E$ subintervals, of size 
$\Delta E = (E_{max}- E_{min})/N_E$.

\item For each fixed value of the wavenumber $k=k_0$, it computes the 
normalized basis functions $C$ and $S$. From these, it computes the function 
$G(E_i)$, at each of the $N_E$ points of the energy interval. These values are 
sent to an array (and stored in a file).
\item The program seeks the roots of the equation $G(E) = 0$, using 
the values obtained previously as a starting point; the energies are 
then refined by a combination of Newton's method (actually, the Secant 
method) and bisection.
\item The resulting roots are stored in a file. Next, move over to the next 
value of $k$ in the $k$-grid ($ k= k_0 + \Delta k$), and repeat the previous step. 
Since this is done for each value of $k= k_j = k_0 + j \Delta k$, one obtains the 
band structure directly from this calculation.

\end{enumerate}

As numerical examples, we show the results obtained for three simple potentials, 
with period $2 \pi$: 

\begin{itemize}

\item The Kronig-Penney model (with $V_0 = 1$, width $a= 0.5$ centered at $x=\pi$)

\begin{eqnarray}
V(x) &=&  0, ~~for~~ |x-\pi|> a \nonumber \\
      &=& V_0 ~~for~~ |x-\pi|<a,
\end{eqnarray}

\item the sinusoidal potential (with $V_0 = 1$ also),
\begin{equation}
V(x) = V_0 ( 1 - cos(x))/2,
\end{equation}

\item and a triangular potential,

\begin{eqnarray}
V(x) &=&  V_0 (\pi - x)/\pi , ~~for~~ x < \pi \nonumber \\
      &=& V_0 (x-\pi)/\pi  ~~for~~ x > \pi .
\end{eqnarray}

\end{itemize}

\begin{figure}[ht]
\includegraphics[width=8.0 cm]{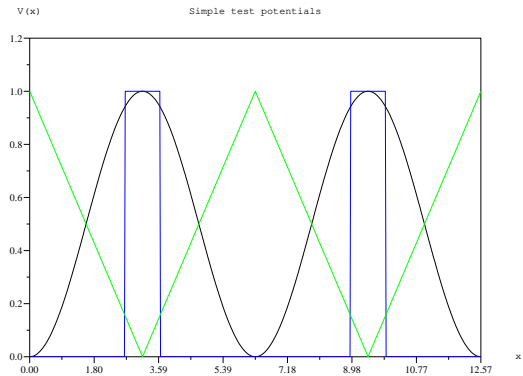}%
\caption{\label{Fig00}The test potentials used in this work: sinusoidal potential $(1-cos(x))/2$, Kronig-Penney
potential centered at $x=\pi$, and a triangular potential.}
\end{figure}

These potentials are plotted together in Figure~\ref{Fig00}, we have chosen them to have the same
maxima, minima, and fundamental period $0 < x < 2 \pi$. These potentials have all been thoroughly
studied in the literature: the Kronig-Penney model has solutions that may be expressed in terms of sinusoidal
and exponential functions. The energy vs. wavenumber relation, $E(k)$, however, is a trascendental 
equation; its solution must be obtained numerically, which somewhat offsets the comparative ease 
with which one obtains the wavefunctions. As described in Stiddard's textbook~\cite{STIDDARD} ( in his
notation: a square barrier, of width $a$, period $a+b$), the eigenvalue equation is 

\begin{equation}
\frac{\gamma^2 - \beta^2}{2 \beta \gamma} sinh(\gamma b) \sin (\beta a) + cosh(\gamma b) \cos(\beta a) 
= cos(k(a + b)).
\end{equation}

\begin{figure}[ht]
\includegraphics[width=8.0 cm]{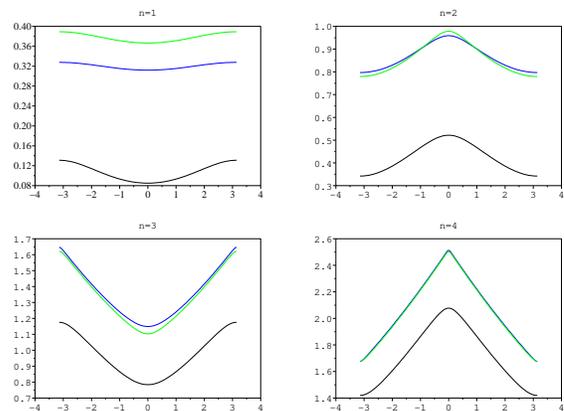}%
\caption{\label{Fig01} First four energy bands for the three model potentials. The Kronig-Penney
potential is shown in the black curve, the sinusoidal potential in the blue curve, and the 
triangular potential on the green curve.}
\end{figure}

The triangular potential may also be solved analitically, in terms of Airy functions, but the 
eigenvalue equation becomes very cumbersome to evaluate. A similar situation occurs for the 
sinusoidal potential, which solutions may be expressed in terms of Mathieu functions. In all cases, 
the analytical calculations needed to obtain the eigenvalue equation from the boundary conditions 
become very involved; much more involved that the numerical solutions described here, and the 'icing on the cake', 
so to speak, is the fact that in the end, all has to be evaluated numerically. So, why not do it numerically
from begining to end?.  The results of our calculations are shown in Figure~\ref{Fig01}.

\section{Final comments}

First, we have established a simple numerical procedure that enables us to compute numerically
the wavefunctions and energy vs. wavevector curve, $E(k)$ for any periodic potential. The procedure 
is very simple, yet it appears that it has not been discussed previously, enabling us to study the 
effect that a parameter change on the potential has on the energies $E(k)$, band gaps and  
densities of states $g(E)$. The procedure may be implemented quite simple using Matlab or Scilab.

\begin{acknowledgments}
The author acknowledges support from Universidad
Austral de Chile (DIDUACH Grant No. S-2004-43), and FONDECYT Grant
No. 1040311. Useful discussion with J.C. Flores and H. Calisto are gratefully acknowledged.
\end{acknowledgments}
\bibliography{bloch}

\begin{thebibliography}{15}
\expandafter\ifx\csname natexlab\endcsname\relax\def\natexlab#1{#1}\fi
\expandafter\ifx\csname bibnamefont\endcsname\relax
  \def\bibnamefont#1{#1}\fi
\expandafter\ifx\csname bibfnamefont\endcsname\relax
  \def\bibfnamefont#1{#1}\fi
\expandafter\ifx\csname citenamefont\endcsname\relax
  \def\citenamefont#1{#1}\fi
\expandafter\ifx\csname url\endcsname\relax
  \def\url#1{\texttt{#1}}\fi
\expandafter\ifx\csname urlprefix\endcsname\relax\def\urlprefix{URL }\fi
\providecommand{\bibinfo}[2]{#2}
\providecommand{\eprint}[2][]{\url{#2}}

\bibitem[{\citenamefont{{Neil Ashcroft and N. David Mermin}}(1976)}]{ASHCROFT}
\bibinfo{author}{\bibnamefont{{Neil Ashcroft and N. David Mermin}}},
  \emph{\bibinfo{title}{Solid State Physics}} (\bibinfo{publisher}{{Holt,
  Rhinehart and Winston}}, \bibinfo{address}{New York, U.S.A.},
  \bibinfo{year}{1976}).

\bibitem[{\citenamefont{{A.A. Bahurmuz and P.D. Loly}}(1981)}]{BAHURMUZ}
\bibinfo{author}{\bibnamefont{{A.A. Bahurmuz and P.D. Loly}}},
  \bibinfo{journal}{Am. J. Phys} \textbf{\bibinfo{volume}{49(7)}},
  \bibinfo{pages}{675} (\bibinfo{year}{1981}).

\bibitem[{\citenamefont{{Sami A. Shakir}}(1983)}]{SHAKIR}
\bibinfo{author}{\bibnamefont{{Sami A. Shakir}}}, \bibinfo{journal}{Am. J.
  Phys.} \textbf{\bibinfo{volume}{59(2)}}, \bibinfo{pages}{845}
  (\bibinfo{year}{1983}).

\bibitem[{\citenamefont{{R. T. Deck and Xiangshan Li}}(1995)}]{DECK}
\bibinfo{author}{\bibnamefont{{R. T. Deck and Xiangshan Li}}},
  \bibinfo{journal}{Am. J. Phys.} \textbf{\bibinfo{volume}{63(10)}},
  \bibinfo{pages}{920} (\bibinfo{year}{1995}).

\bibitem[{\citenamefont{{S.A. Denham, B.C. Harms and S. T.
  Jones}}(1982)}]{DENHAM}
\bibinfo{author}{\bibnamefont{{S.A. Denham, B.C. Harms and S. T. Jones}}},
  \bibinfo{journal}{Am. J. Phys.} \textbf{\bibinfo{volume}{50(4)}},
  \bibinfo{pages}{374} (\bibinfo{year}{1982}).

\bibitem[{\citenamefont{{Juan F. Van der Maelen Ur\'{\i}a, Santiago
  Garc\'{\i}a-Granda and Amador Men\'endez-Vel\'asquez}}(1996)}]{VDMAELEN}
\bibinfo{author}{\bibnamefont{{Juan F. Van der Maelen Ur\'{\i}a, Santiago
  Garc\'{\i}a-Granda and Amador Men\'endez-Vel\'asquez}}},
  \bibinfo{journal}{Am. J. Phys.} \textbf{\bibinfo{volume}{64(3)}},
  \bibinfo{pages}{327} (\bibinfo{year}{1996}).

\bibitem[{\citenamefont{Leung}(1993)}]{LEUNG}
\bibinfo{author}{\bibfnamefont{K.}~\bibnamefont{Leung}}, \bibinfo{journal}{Am.
  J. Phys.} \textbf{\bibinfo{volume}{61(11)}}, \bibinfo{pages}{1020}
  (\bibinfo{year}{1993}).

\bibitem[{\citenamefont{{B. M\'endez and F.
  Dom\'{\i}nguez-Adame}}(1994)}]{MENDEZ}
\bibinfo{author}{\bibnamefont{{B. M\'endez and F. Dom\'{\i}nguez-Adame}}},
  \bibinfo{journal}{Am. J. Phys.} \textbf{\bibinfo{volume}{62(2)}},
  \bibinfo{pages}{143} (\bibinfo{year}{1994}).

\bibitem[{\citenamefont{{R. C. Greenhow and J. A. D. Mathew}}(1992)}]{GREENHOW}
\bibinfo{author}{\bibnamefont{{R. C. Greenhow and J. A. D. Mathew}}},
  \bibinfo{journal}{Am. J. Phys.} \textbf{\bibinfo{volume}{60(7)}},
  \bibinfo{pages}{655} (\bibinfo{year}{1992}).

\bibitem[{\citenamefont{Johnston}(1992)}]{JOHNSTON}
\bibinfo{author}{\bibfnamefont{I.~D.} \bibnamefont{Johnston}},
  \bibinfo{journal}{Am. J. Phys.} \textbf{\bibinfo{volume}{60(7)}},
  \bibinfo{pages}{600} (\bibinfo{year}{1992}).

\bibitem[{\citenamefont{{P. J. Cooney, E. P. Kanter and Z.
  Vager}}(1981)}]{COONEY}
\bibinfo{author}{\bibnamefont{{P. J. Cooney, E. P. Kanter and Z. Vager}}},
  \bibinfo{journal}{Am. J. Phys.} \textbf{\bibinfo{volume}{49(1)}},
  \bibinfo{pages}{76} (\bibinfo{year}{1981}).

\bibitem[{\citenamefont{{Debra J. Searles and Ellak I. von
  Nagy-Felsobuki}}(1988)}]{SEARLES}
\bibinfo{author}{\bibnamefont{{Debra J. Searles and Ellak I. von
  Nagy-Felsobuki}}}, \bibinfo{journal}{Am. J. Phys.}
  \textbf{\bibinfo{volume}{56(5)}}, \bibinfo{pages}{444}
  (\bibinfo{year}{1988}).

\bibitem[{\citenamefont{{P. C. Chow}}(1972)}]{CHOW}
\bibinfo{author}{\bibnamefont{{P. C. Chow}}}, \bibinfo{journal}{Am. J. Phys.}
  \textbf{\bibinfo{volume}{40)}}, \bibinfo{pages}{730} (\bibinfo{year}{1972}).

\bibitem[{\citenamefont{{W. H. Press, B. P. Flannery, S. A. Teukolsky and W. P.
  Vetterling}}(1989)}]{RECIPES}
\bibinfo{author}{\bibnamefont{{W. H. Press, B. P. Flannery, S. A. Teukolsky and
  W. P. Vetterling}}}, \emph{\bibinfo{title}{Numerical Recipes in Pascal}}
  (\bibinfo{publisher}{{Cambridge University Press}}, \bibinfo{address}{New
  York, U.S.A.}, \bibinfo{year}{1989}).

\bibitem[{\citenamefont{{M. H. B. Stiddard}}(1975)}]{STIDDARD}
\bibinfo{author}{\bibnamefont{{M. H. B. Stiddard}}}, \emph{\bibinfo{title}{The
  Elementary Language of Solid State Physics}} (\bibinfo{publisher}{{Academic
  Press}}, \bibinfo{address}{New York, U.S.A.}, \bibinfo{year}{1975}).

\end{thebibliography}
\end{document}